\documentclass[11pt,twoside]{article}
\usepackage{./asp2014}
\usepackage{color}

\hypersetup{
 colorlinks= true, 
 urlcolor  = black, 
 linkcolor = red, 
 citecolor = blue 
} 

\aspSuppressVolSlug
\resetcounters

\begin{document}

\title{Precision Gas-dynamical Mass Measurement of Supermassive Black Holes with the ngVLA}
\author{Benjamin D. Boizelle$^1$, Kristina Nyland$^2$, and Timothy A. Davis$^3$\\
\affil{$^1$George P.\,and Cynthia Woods Mitchell Institute for Fundamental Physics and Astronomy, Department of Physics and Astronomy, Texas A\&M University, College Station, TX 77843; \email{bboizelle@tamu.edu}}
\affil{$^2$National Radio Astronomy Observatory, Charlottesville, VA, 22903; \email{knyland@nrao.edu}}
\affil{$^3$School of Physics \& Astronomy, Cardiff University, Queens Buildings, The Parade, Cardiff, Wales, UK\\ \email{DavisT@cardiff.ac.uk}}
}

\paperauthor{Benjamin D. Boizelle}{bboizelle@physics.tamu.edu}{}{Department of Physics and Astronomy, Texas A\&M University}{}{College Station}{TX}{77843}{USA}
\paperauthor{Kristina Nyland}{knyland@nrao.edu}{}{National Radio Astronomy Observatory}{}{Charlottesville}{VA}{22903}{USA}
\paperauthor{Timothy A. Davis}{DavisT@cardiff.ac.uk}{}{Department of Physics and Astronomy, Cardiff University}{}{Cardiff}{Wales}{CF24 3AA}{UK}

\begin{abstract}
Emission line observations of circumnuclear gas disks in the ALMA era have begun to resolve molecular gas tracer kinematics near supermassive black holes (BHs), enabling highly precise mass determination in the best cases. The ngVLA is capable of extremely high spatial resolution imaging of the CO(1$-$0) transition at 115 GHz for nearby galaxies. Furthermore, its high (anticipated) emission line sensitivity suggests this array can produce benchmark BH mass measurements. We discuss lessons learned from gas-dynamical modeling of recent ALMA data sets and also compare ALMA and ngVLA CO simulations of a dynamically cold disk. While only a fraction of all local galaxies likely possess sufficiently bright, regularly-rotating nuclear molecular gas, in such cases the ngVLA is expected to more efficiently resolve such emission arising at a projected 50$-$100 mas from the central BH.
\end{abstract}


\section{Introduction}

The observed properties of massive, bulge-dominated galaxies support the existence of a strong co-evolution between the formation and growth of their stellar spheroids and the supermassive black holes (BHs) residing in their nuclei (e.g., \citealt{kormendy+13, heckman+14}).  Accurate BH mass measurements are necessary to investigate BH growth over cosmic time as well as to improve constraints on theoretical models of their formation at high redshift.  Despite the importance of these measurements to our understanding of galaxy evolution, the full distribution of BH masses at fixed galaxy properties such as morphology, redshift, and stellar mass -- particularly for lower-mass central BHs -- remains poorly constrained (e.g., \citealt{greene+16}). ALMA is now beginning to have its long-anticipated \citep[e.g.,][]{maiolino+08} revolutionary impact on circumnuclear kinematic studies, already facilitating precision BH mass measurements \citep[e.g.,][]{barth+16a,davis+17} using more common molecular gas tracers. In this science case, we first describe the benefits and challenges associated with cold molecular gas emission as a kinematic tracer, and afterwards detail the promise of the ngVLA to enable even more sensitive, higher angular resolution exploration of the central gravitational potentials of nearby galaxies.

\begin{figure*}[t!]
\centering
\includegraphics[clip=true, trim=0cm 0cm 0cm 0cm, height=0.65\textwidth]{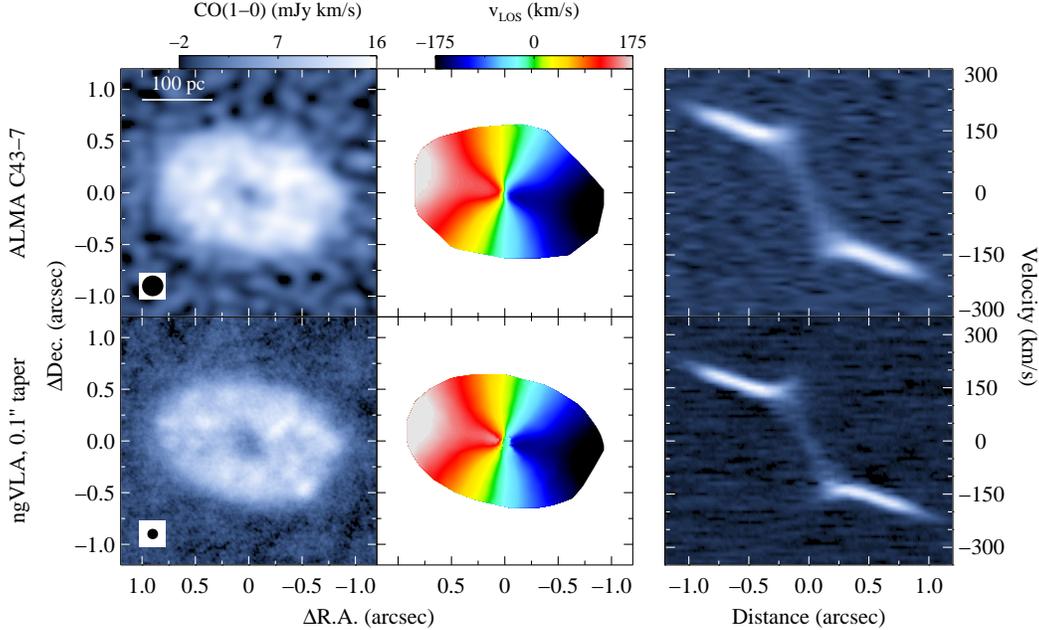}
\caption{Line moment maps and position-velocity diagrams taken from full-cube ALMA 0.2$^{\prime \prime}$ (\textit{top row}) and ngVLA 0.1$^{\prime \prime}$ (\textit{bottom row}; after applying an outer $uv$ taper) simulations of CO(1$-$0) disk rotation. The randomly generated clumpy circumnuclear disk at a distance of $\sim30$ Mpc has a central hole (with radius $\sim0.2^{\prime \prime}\approx30$ pc) in CO-bright gas and a plausible host galaxy gravitational potential \citep[similar to the case found in][]{davis+18}, with a BH mass of $2\times10^8$ $M_\odot$ (and $r_{\rm g}\sim0.3^{\prime \prime}$). The same 2 hr on-source integration time reaches 0.46 and 0.14 mJy/beam rms sensitivities in 10 km s$^{-1}$ channels for the ALMA and ngVLA data sets, respectively. Simulations show that the ngVLA enables better identification and isolation of the faint central CO emission. This primary kinematic signature, buried in higher line moments of the ALMA model cube, becomes more distinct in the simulated ngVLA line-of-sight centroid velocity field. Even in cases where the molecular gas does not originate very deep within $r_{\rm g}$, ngVLA observations will promote more confident $M_{\rm BH}$ determination \citep[e.g., see][]{barth+16a,barth+16b}. When the CO emission tracers Keplerian rotation, high angular resolution imaging with this proposed array will enable benchmark BH mass measurements.}
\label{fig1}
\end{figure*}

\section{Measuring BH masses with Molecular Gas Tracers}
Gas-dynamical modeling of a thin coplanar disk within a galaxy nucleus is an appealing technique to determine its central BH mass ($M_{\rm BH}$). In addition to conceptual and computational simplicity, idealized rotating disks are also free from most of the potentially serious systematics that affect stellar-dynamical $M_{\rm BH}$ measurements \citep[e.g.,][]{van10,mcconnell+13b} and can be applied to systems with larger-scale stellar non-axisymmetries (such as bars and mergers). Many early studies used observations of ionized atomic gas disks to map the gravitational potentials of galaxy nuclei, but modeling complications \citep[especially gas turbulence; see][]{van98,wal10} limit the confidence of $M_{\rm BH}$ determinations from such kinematically warm tracers \citep{kormendy+13}. Dynamically cold molecular emission is a promising avenue as such relaxed systems retain the benefits of disk-like rotation while avoiding the primary systematics afflicting the treatment of ionized gas disks.  In most cases, these dynamically cold disks are found to be nearly ideal central probes of nearby galaxies. The centers of many late and early-type galaxies (LTGs/ETGs), possessing both active and quiescent nuclei, contain significant molecular gas reservoirs \citep[$\gtrsim10^7$ $M_\odot$;][]{bol17,alatalo+13}, indicating that BH masses in nearly all types of galaxies may be measured in the same manner by observing and modeling cold molecular gas rotation.

Unambiguously detecting a BH's kinematic impact hinges on (at least approximately) resolving its sphere of influence $r_{\rm g}\approx G M_{\rm BH}/\sigma_\star^2$ wherein $M_{\rm BH}$ dominates over the extended mass contributions and gives rise to an elevated stellar velocity dispersion $\sigma_\star$. Even moderate ($0.3^{\prime \prime}-0.5^{\prime \prime}$) resolution, sensitive molecular line observations with ALMA have the potential to provide dynamical BH mass measurements (or strong constraints) for perhaps several hundred nearby ($\lesssim100$ Mpc) galaxies \citep{davis+14}, thereby greatly increasing the number of $M_{\rm BH}$ determinations made through dynamical means. Precision BH mass measurements (at the few percent level or better) require resolving kinematic tracers well within $r_{\rm g}$, where the observed rotation is maximally sensitive to the BH influence. In the best (but rare) examples, extremely well-resolved extragalactic H$_2$O megamaser disks trace out clear Keplerian rotation and yield ``gold standard'' $M_{\rm BH}$ values (generally $\sim10^7$ $M_\odot$) for these active galaxies \citep[e.g.,][]{miyoshi+95,kuo+11}.

Disentangling the central rapid gas velocities of more common molecular tracers such as CO has been largely unachievable by the previous generation of mm/sub-mm arrays due to beam smearing: for example, a $10^8$ $M_\odot$ BH will have a projected $r_{\rm g}\sim0.2^{\prime \prime}$ when viewed at a distance of 20 Mpc. For one such nearby galaxy, \citet{davis+13b} used the CARMA array to just barely resolve CO(2$-$1) emission originating within the sphere of influence and permit the first BH mass constraint using CO kinematics. Now, ALMA facilitates much deeper, higher angular resolution imaging of this and other tracers in a fraction of the time. Data sets that just marginally resolve $r_{\rm g}$ have already produced gas-dynamical BH mass measurements with total uncertainties \citep[on the order of $20-30\%$, including systematics;][]{barth+16b,onishi+17,davis+17,davis+18} that are commensurate with those derived from stellar-dynamical modeling. Observations that more fully resolve $r_{\rm g}$ can produce benchmark $M_{\rm BH}$, enabling tests on the assumptions used in other techniques when applied to measure the same BH mass \citep[e.g.,][]{barth+16a,rusli+11}.


\section{Lessons From ALMA Observations}
Over an order of magnitude higher maximum angular resolution imaging at $\sim115$ GHz than ALMA initially suggests that the ngVLA will make possible correspondingly more precise BH mass measurements. This array is theoretically capable of resolving the spheres of influence for a Milky Way-analog BH observed at $\sim100$ Mpc, as well as for $\sim40\%$ of the galaxies in SDSS. However, not all nearby galaxies possess bright circumnuclear molecular gas emission originating from within $r_{\rm g}$. Among those that do, some fraction will exhibit relaxed central gas kinematics. Given the relatively smaller angular $r_{\rm g}$ sizes of LTGs, kinematic studies that resolve $r_{\rm g}$ for these targets are still nascent. We here focus on ETGs as they provide a clearer illustration. Roughly half of all luminous E/S0 hosts harbor detectable CO emission, although perhaps in only $\sim10\%$ does the gas exhibit dynamically cold rotation \citep{alatalo+13}. Recent ALMA observations reveal that commonly utilized molecular species tend to be dissociated (or are at least underluminous) in the central 10-100 pc \citep{onishi+17,barth+16a,boizelle+17,davis+18}.

\citet{boizelle+17} estimate that perhaps a few percent of all ETGs will be prime candidates for precision BH mass measurement using ALMA. A handful of these (and some unknown fraction of LTGs) may still show strong, unresolved Keplerian-like rotation at the edge of ALMA capabilities, warranting even higher angular resolution line imaging with the ngVLA. Obtaining such well-resolved observations, with several independent resolution elements across $r_{\rm g}$, alleviates nearly all of the remaining degeneracies \citep[e.g., between $M_{\rm BH}$ and the gas turbulent velocity dispersion;][]{barth+16a,barth+16b} and systematics \citep[e.g., uncertainty in the extended mass velocity profile due to central dust obscuration;][Boizelle et al.\ 2018, in prep.]{barth+16b}. We believe this proposed cm/mm array will be optimally utilized when following up on previous interferometric observations (with $\theta_{\rm 1/2}\lesssim0.25^{\prime \prime}$) that clearly identify gas emission within $r_{\rm g}$. Its anticipated sensitivity will also aid in detecting faint central emission that is otherwise beneath the limit of previous ALMA observations.

\begin{figure*}[t!]
\centering
\includegraphics[clip=true, trim=0cm 0cm 0cm 0cm, height=0.65\textwidth]{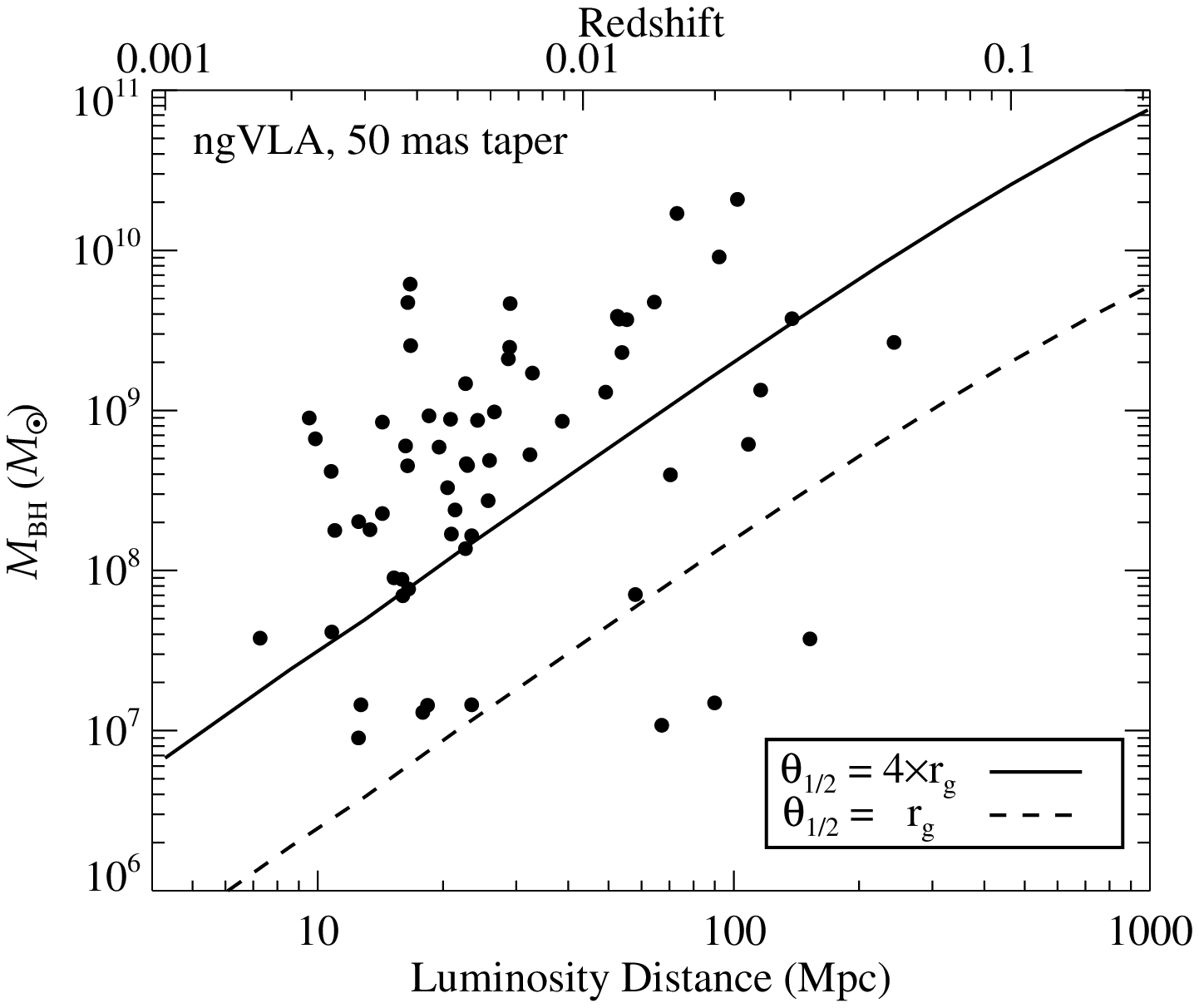}
\caption{Minimum BH mass detectable using the ngVLA at $\sim115$ GHz (with a \textit{uv} taper to achieve 50 mas angular resolution) as a function of luminosity distance assuming a standard cosmology \citep{Planck15}. Data points represent the elliptical and classical bulge galaxies on which the $M_{\mathrm{BH}}-\sigma_{*}$ relation is currently based \citep{kormendy+13}. Curves show the mass limit reachable by the ngVLA using CO(1$-$0) observations that just marginally resolve the BH sphere of influence (\textit{dashed}; as is typical for most stellar/gaseous $M_{\rm BH}$ determinations) and that more highly resolve $r_{\rm g}$ (\textit{solid}); assuming detectable CO emission in regular rotation within $r_{\rm g}$, expected total mass uncertainties are about $20-30\%$ and $\lesssim10\%$, respectively.}
\label{fig2}
\end{figure*}

\section{ngVLA Unique Capabilities}

High-resolution observations of the CO(1$-$0) transition (with line excitation temperature $T_{\rm ex}\sim5-10$ K) using ALMA are limited by extended configuration overheads and brightness temperature sensitivities. Typical CO surface brightness measurements suggest that ALMA is capable of resolving CO(1$-$0) emission interior to (with about four independent resolution elements across) $r_{\rm g}$ for nearby galaxies with large projected spheres of influence \citep[$\sim0.2^{\prime \prime}-0.3^{\prime \prime}$; e.g.,][]{boizelle+17,davis+17,davis+18}. An integrated line S/N$\gtrsim$10 ensures confident detection of kinematic features in these disks with low intrinsic line widths \citep[$\sigma\sim5-20$ km s$^{-1}$; see][]{barth+16a}. Achieving this S/N level requires rms sensitivities $\sim0.2-0.4$ mJy/beam in 10 km s$^{-1}$ channels, with the resulting ALMA 115 GHz on-source (total) integrations times ranging between $3-14$ ($9-32$) hours. As a result of these requirements, kinematic studies of local galaxies with ALMA often focus on the 2$-$1 transition to better balance limiting sensitivities, weather constraints, and achievable angular resolution. Based on current performance metrics, ngVLA CO(1$-$0) observations with the same channel spacing are expected to reach of order $\sim100$ microJy/beam line rms (producing S/N$\sim$20) in a single hour (on-source) on 100 mas scales. These expected results also improve on ALMA Band 6 forecasts, more than halving the time allocation to accurately measure CO line properties at $\sim0.1^{\prime \prime}$ resolution.

In Figure~\ref{fig1}, we show our simulated ALMA and ngVLA full-cube CO(1$-$0) observations of a model molecular disk to compare array capabilities. The combination of higher angular resolution and lower limiting sensitivities enable detection of additional faint, high-velocity central emission that promotes a more confident $M_{\rm BH}$ determination. Slightly higher angular resolution ngVLA imaging of such circumnuclear disks should be possible on 50 mas scales for reasonable (a few hours) integration times. On these angular scales, moderately accurate ($\sim20\%$) $M_{\rm BH}$ determinations of $\sim10^9$ $M_\odot$ should be possible for galaxies at distances of up to 300 Mpc, with precision ($\lesssim10\%$) measurements to nearly 75 Mpc (see Figure~\ref{fig2}). Almost two orders of magnitude higher resolution CO(1$-$0) observations are \textit{technically} feasible with this cm/mm array. However, the typical CO emission deficit at radii deep within $r_{\rm g}$ and prohibitively long integration times suggest that ngVLA extreme angular resolution observations are not likely to be very useful for BH studies in most cases.

The above scenario focused on galaxies with large angular $r_{\rm g}$ sizes, tending to favor nearby ETGs. The anticipated array performance suggests that ngVLA CO imaging is likely able to identify lower mass ($\sim10^7$ $M_\odot$) BH kinematic signatures in late-type targets out to at least 20 Mpc. High-resolution cold molecular gas observations of such (often dusty) nuclei will result in more confident mass determinations than typically obtained using ionized atomic tracers \citep{beifiori+09} as well as better constraints on the low-mass end of the BH/host galaxy scaling relationships. This proposed array is also likely to be impactful in extragalactic H$_2$O megamaser BH mass measurements in active galaxies that provide direct geometrical measurement of $H_0$ (see Braatz et al.\,2018, this work). While its longest baselines are not able to map out the 22 GHz emission at sub-mas resolution, ngVLA will provide an important large-collecting area element in new and revised VLBI campaigns at greater sensitivity.


\section{Multiwavelength Synergy}
The next generation of giant segmented mirror telescopes (such as the E-ELT and the TMT) will offer angular resolutions down to mas scales, comparable to those technically achievable using the ngVLA in its high-frequency bands. These optical/NIR observatories will dramatically increase the number of unobscured nuclei for which it is possible to directly probe stellar and gaseous kinematics within the BH sphere of influence. Space-based NIR observations with the \textit{James Webb Space Telescope} are expected to expand the sample of active galaxies with quality $M_{\rm BH}$ values using techniques such as reverberation mapping. Ongoing ALMA work is now providing tight BH mass constraints for galaxies with obscured nuclei, and the ngVLA contributions are expected to increase both the precision and number of these dynamical measurements. These highly complimentary avenues promise to more fully fill in the BH census with high-quality measurements spanning BH mass and host galaxy properties. Increasing the number of candidates for detailed dynamical modeling will similarly increase the number of cases where multiple techniques are applicable to the same target. Such precision $M_{\rm BH}$ values will enable crucial studies of the underlying assumptions and systematics inherent to each method.


\section{Summary}
The ngVLA will expand on ALMA's revolutionary capability to probe cold molecular gas rotation within BH $r_{\rm g}$. Anticipated factor of at least $\sim2-4$ improvements in line sensitivity with the ngVLA should enable detection of additional faint circumnuclear CO(1$-$0) emission, and an estimated maximum angular resolution of about 50 mas will better resolve Keplerian rotation for galaxies out to a greater distance. The ngVLA is therefore expected to drive the field of BH mass measurement towards more highly resolving cold gas kinematics that in turn facilitate more precise $M_{\rm BH}$ determinations.

\bibliography{ms}  

\end{document}